\documentclass[groupaddress,9pt,twocolumn,superscriptaddress, aps, prl]{revtex4-1}
\usepackage{lmodern}
\usepackage{textcomp}
\usepackage{graphicx}
\usepackage{amsmath}
\usepackage{graphicx}
\usepackage{bm}
\usepackage{array}
\usepackage{dcolumn}
\usepackage{hepparticles}
\usepackage{heppennames}
\usepackage{hepnicenames}
\usepackage{hyperref}
\usepackage{lineno}
\usepackage[english]{babel}

\newcommand{\ket}[1]{|#1\rangle}
\newcommand{\bra}[1]{\langle #1|}
\newcommand{\Yb}{$^{171}{\rm{Yb}}^{+} $}

\renewcommand{\thefigure}{\textbf{\arabic{figure}}} 
\newcommand{\FigureOne}{
\begin{figure}
\begin{center}
\includegraphics[width = 87 mm]{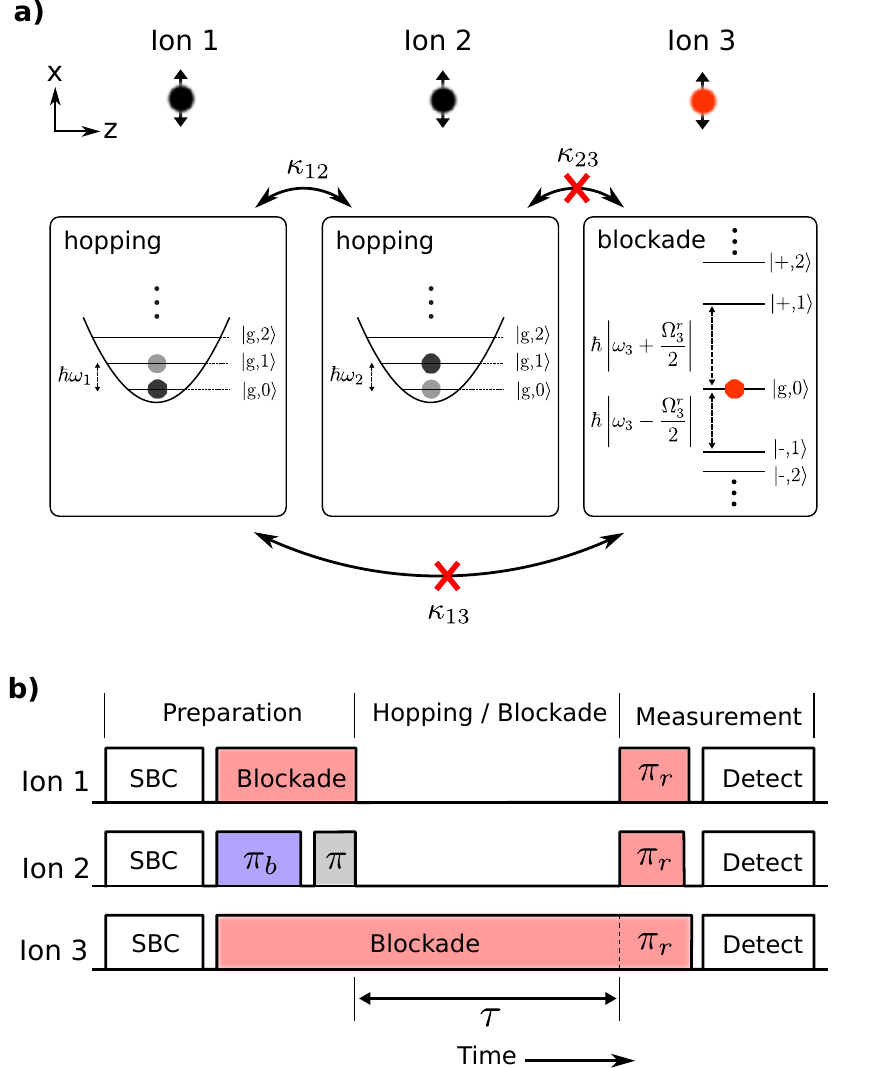}
\end{center}
\renewcommand{\baselinestretch}{1}
\small\normalsize
\caption{ Experimental system for observing hopping of a single phonon excitation between local transverse motional modes along the X-direction. (a) The local phonon frequencies are represented by $\omega_i$ in a frame rotating at the transverse common mode frequency $\omega_x$, and $\kappa_{jk}$ is the phonon hopping strength between modes $j$ and $k$. Phonon blockades on individual sites (here ion 3) is implemented by driving resonant red sideband transitions with strength $\Omega^r_j$ that gives rise to an energy splitting between the ground state $\ket{g,0}$ and the first excited polaritonic states $\ket{\pm,1}$. (b) An experimental sequence where each ion is prepared in the ground state of spin and motion $\ket{g,0}$ using Raman sideband cooling (SBC). A single phonon is excited on ion $2$ using $\pi$-pulses at the blue sideband ($\pi_b$) and carrier ($\pi$) transitions. Local phonon blockades are applied using resonant red sideband pulses (shown in red). The hopping duration $\tau$  is varied to observe the dynamics of local phonon occupancy ($0$ or $1$ phonon) measured by first projecting it to the internal spin states ($\ket{g}$ or $\ket{e}$) of each ion using red sideband $\pi$-pulses ($\pi_r$) followed by the detection of state-dependent fluorescence from each ion using a photomultiplier tube array.}
\label{Fig1}
\end{figure}}

\newcommand{\FigureTwo}{
\begin{figure*}
\begin{center}
\includegraphics[width = 183 mm]{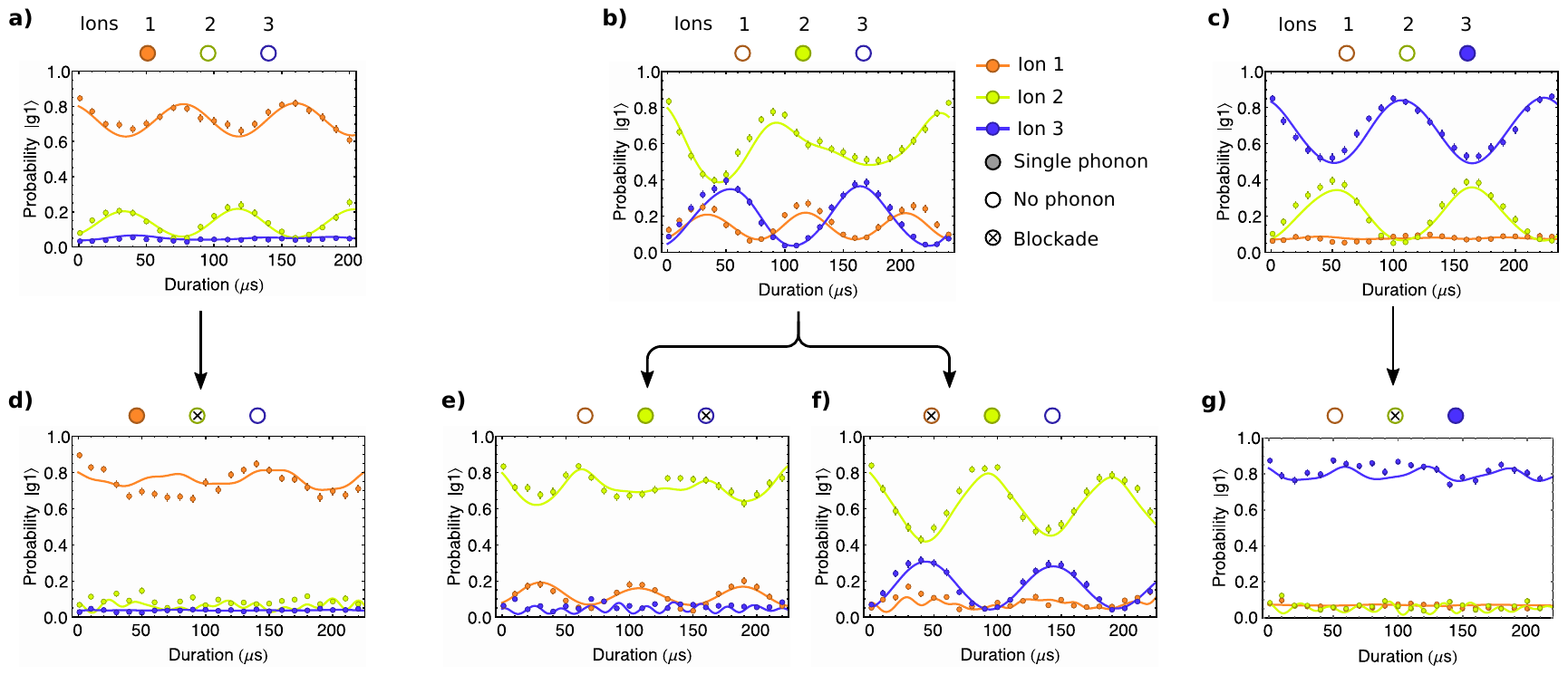}
\end{center}
\renewcommand{\baselinestretch}{1}
\small\normalsize
\caption{ The evolution of local phonon occupancies with initial single-phonon excitations on ions 1, 2, and 3 as shown by the shaded orange, green, and blue circles, respectively. In the absence of a blockade (a-c), the dynamics are governed by the hopping strengths $\{\kappa_{jk}\}$ and the local mode frequencies $\{\omega_j\}$. The corresponding dynamics in the presence of a blockade (d-g) indicate hopping suppression, which is determined by the  blockade strength $\{\Omega^r_j\}$. The theoretical plots are obtained by fitting a Jaynes-Cummings Hubbard model (Hamiltonian in Eq. \ref{Eq1} and \ref{Eq2}) with free parameters $\{\Omega^r_j\}$, $\{\omega_j\}$ and $\{\kappa_{jk}\}$ using all evolution data sets collectively. Error bars represent statistical uncertainties of $2\sigma$.}
\label{Fig2}
\end{figure*}}

\newcommand{\FigureThree}{
\begin{figure}
\begin{center}
\includegraphics[width = 85 mm]{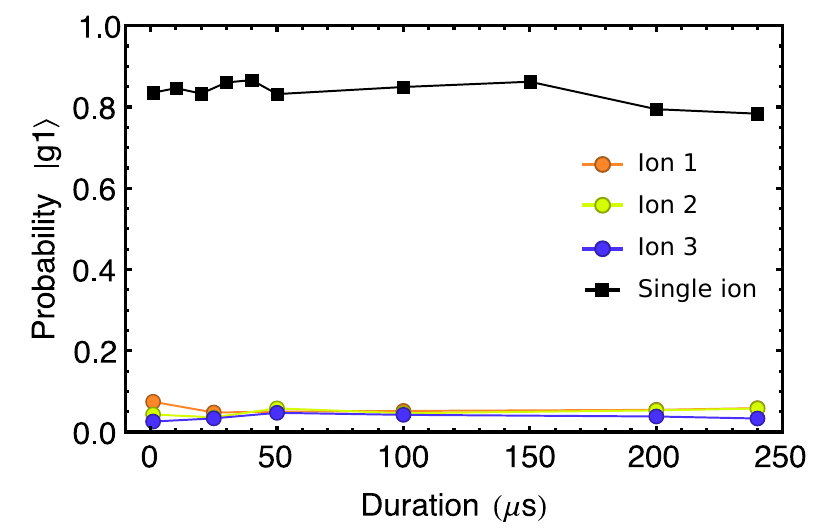}
\end{center}
\renewcommand{\baselinestretch}{1}
\small\normalsize
\caption{Steady phonon occupancy observed in the absence of hopping when a single ion is prepared in state $\ket{g,1}$ or in a chain of three ions where each is sideband cooled and prepared in state $\ket{g,0}$. For the single ion experiment, a constant phonon occupancy indicates negligible crosstalk with other transverse and axial modes of the ion trap. In a three ion chain, the near-zero phonon occupancy indicates that the combined effect of the local mode heating rate and crosstalk with other transverse modes (that are not sideband cooled) is negligible. The duration of both experiments is similar to that used to observe the hopping dynamics in Fig. \ref{Fig2}. The probabilities of detecting single phonon excitations in both cases are used to estimate the state preparation and measurement (SPAM) fidelity of states $\ket{g,0}$ and $\ket{g,1}$. Error bars showing statistical uncertainties of $2\sigma$ are smaller than experimental data points.
}
\label{Fig3}
\end{figure}}

\newcommand{\TableOne}{
\begin{table}
\begin{center}
\begin{tabular} { c c c }
Parameter\hspace{2 mm} & Fitted value\hspace{2 mm} & Measured value\\[0.5ex]
\hline
\\
 $\omega_{12}$ & $11.58\:$ & --- \\ [0.5ex]
 $\omega_{23} $ & $7.36\:$ & ---  \\  [0.5ex]
 $\kappa_{12}$ & $2.90$\: & $3.27(19)\:$   \\ [0.5ex]
 $\kappa_{23}$ & $2.96\:$ & $3.36(20)\:$\\ [1ex]
 $\Omega_1^r$ & $39.7\:$& $43.1(16)\:$\\[1ex]
 $\Omega_2^r$ & $45.9\:$& $47.6(14)\:$\\[1ex]
 $\Omega_3^r$ & $46.3\:$& $46.0(19)\:$\\[1ex] 
 \hline
\end{tabular}
\end{center}
\caption{ Observed experimental parameters relevant to hopping and the blockade in units of $ 2\pi\times$kHz. The values obtained from fits to the hopping data (Fig.\ref{Fig2}) are compared with those obtained from direct measurement. The measured hopping rate $\kappa_{ij}$ is obtained from inter-ion distances $\{d_{12}, d_{23}\}=\{10.1(2),10.0(2)\}\:\mu$m, where the systematic error is due to uncertainty in $d_{ij}$. The measured red-sideband Rabi frequency is directly obtained from sideband spectroscopy (see Fig.\ref{FigS1}). The local mode frequencies measured from sideband spectroscopy are not given due to large Stark shifts that vary between experimental runs with beam alignment \cite{Lee2016}.}
\label{Tab1}
\end{table}
}

\newcommand{\FigureFour}{
\begin{figure*}
\begin{center}
\includegraphics[width = 183 mm]{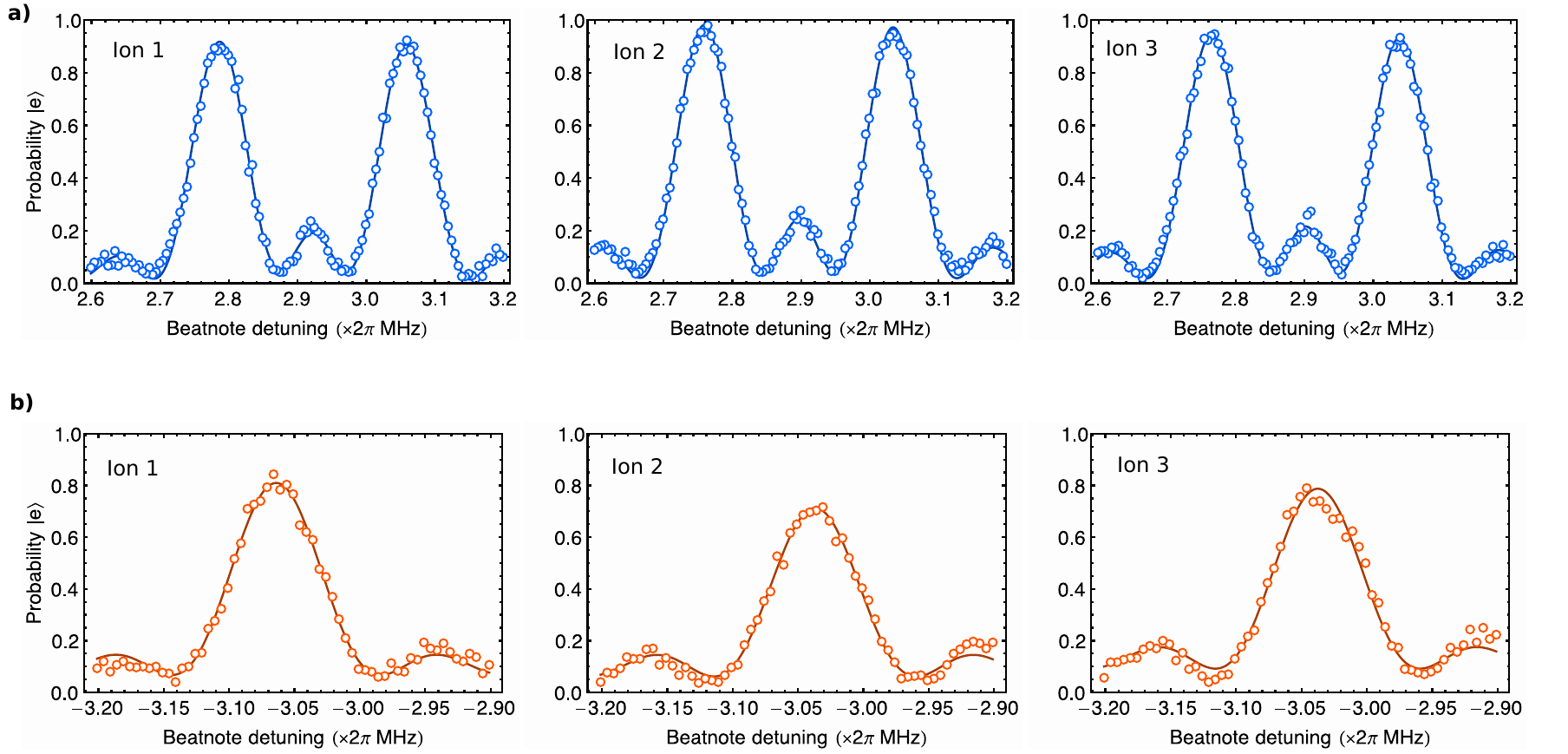}
\end{center}
\renewcommand{\baselinestretch}{1}
\small\normalsize
\caption{ Sideband spectroscopy of local phonon modes. (a) The blue sideband spectroscopy showing sideband transitions for the two transverse local modes of each ion in a chain of three. Each ion is Raman sideband-cooled to the motional ground state of both modes. A blue sideband $\pi-$pulse is then applied on each ion, which changes its state from $\ket{g,0}$ to $\ket{e,1}$ when resonant with a sideband transition. The two transverse local modes are sufficiently separated by $\approx 250\:$kHz compared to the sideband transition strength such that the mode with higher energy can be spectrally addressed for introducing local phonon excitations during the `preparation' step of an experimental sequence as shown in Fig. \ref{Fig1}b. (b) Red sideband spectroscopy of the higher energy transverse local mode. Each ion is initialized to $\ket{g,1}$ as shown in Fig. \ref{Fig1}b followed by a red sideband $\pi-$pulse that flips the state from $\ket{g,1}$ to $\ket{e,0}$ followed by measurement of the spin-up state $\ket{e}$ using state-dependent fluorescence. The red sideband Rabi frequencies obtained by fitting the data determines the strength of phonon blockade applied on each ion (see Fig. \ref{Fig1}a).}
\label{FigS1}
\end{figure*}}

\newcommand{\FigureFive}{
\begin{figure*}
\begin{center}
\includegraphics[width = 183 mm]{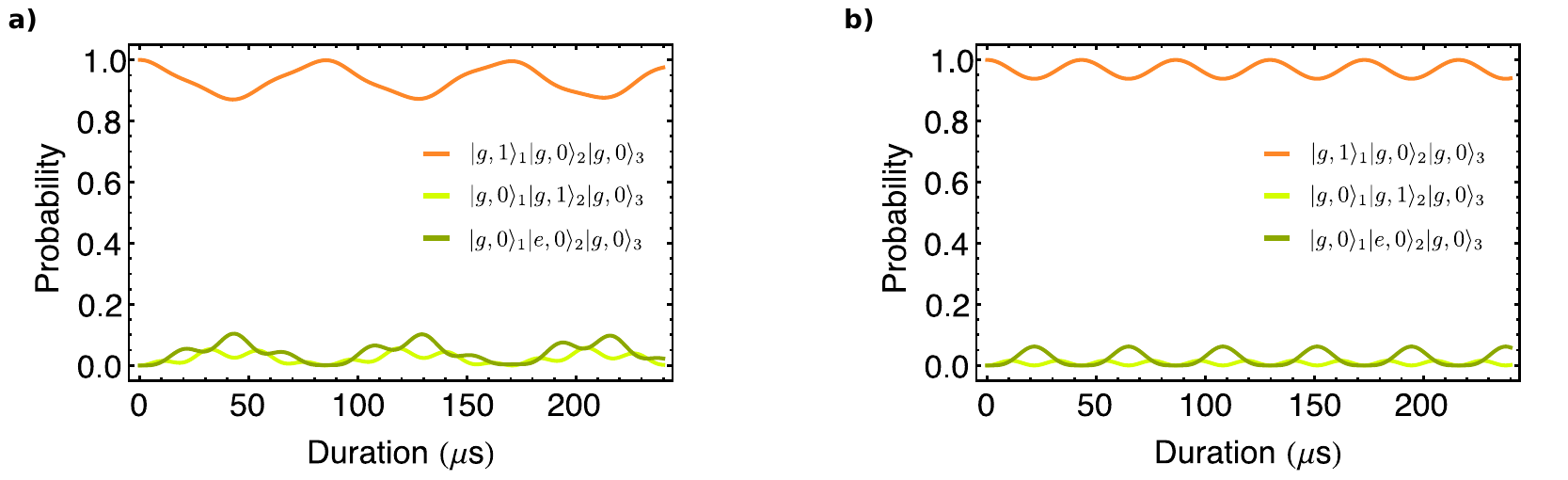}
\end{center}
\renewcommand{\baselinestretch}{1}
\small\normalsize
\caption{ A theoretical analysis of the effect of phonon blockade on ion 2 when an initial phonon excitation is introduced on ion 1. The analysis captures phonon hopping in terms of the time evolution of the states $\ket{g,1}_1\ket{g,0}_2\ket{g,0}_3$, $\ket{g,0}_1\ket{g,1}_2\ket{g,0}_3$, and $\ket{g,0}_1\ket{e,0}_2\ket{g,0}_3$, which are predominantly coupled to each other via the hopping and blockade interactions. (a) Shows the dynamics for real experimental conditions where $\omega_{12}/2\pi=11.58\:$kHz. This simulates experimental data in Fig. \ref{Fig2}d. (b) Shows the dynamics for the case where $\omega_{12}/2\pi=0\:$kHz which exhibits a higher suppression of hopping.}
\label{FigS2}
\end{figure*}}

\begin{abstract}
The local phonon modes in a Coulomb crystal of trapped ions can represent a Hubbard system of coupled bosons. We selectively prepare single excitations at each site and observe free hopping of a boson between sites, mediated by the long-range Coulomb interaction between ions. We then implement phonon blockades on targeted sites by driving a Jaynes-Cummings interaction on individually addressed ions to couple their internal spin to the local phonon mode. The resulting dressed states have energy splittings that can be tuned to suppress phonon hopping into the site. This new experimental approach opens up the possibility of realizing large-scale Hubbard systems from the bottom up with tunable interactions at the single-site level. 
\end{abstract}

\begin{document}

\title{Observation of Hopping and Blockade of Bosons in a Trapped Ion Spin Chain}
\author{S. Debnath}\email{sdebnath@berkeley.edu}
\affiliation{Joint Quantum Institute, Department of Physics, and Joint Center for Quantum Information and Computer Science, University of Maryland, College Park, MD 20742, USA}
\author{N. M. Linke} 
\affiliation{Joint Quantum Institute, Department of Physics, and Joint Center for Quantum Information and Computer Science, University of Maryland, College Park, MD 20742, USA}
\author{S.-T. Wang}
\affiliation{Department of Physics, University of Michigan, Ann Arbor, Michigan 48109, USA}
\author{C. Figgatt}
\affiliation{Joint Quantum Institute, Department of Physics, and Joint Center for Quantum Information and Computer Science, University of Maryland, College Park, MD 20742, USA}
\author{K. A. Landsman}
\affiliation{Joint Quantum Institute, Department of Physics, and Joint Center for Quantum Information and Computer Science, University of Maryland, College Park, MD 20742, USA}
\author{L.-M. Duan}
\affiliation{Department of Physics, University of Michigan, Ann Arbor, Michigan 48109, USA}
\author{C. Monroe}
\affiliation{Joint Quantum Institute, Department of Physics, and Joint Center for Quantum Information and Computer Science, University of Maryland, College Park, MD 20742, USA}

\maketitle

Trapped atomic ions are an excellent medium for quantum computation and quantum simulation, acting as a many-body system of spins with programmable and reconfigurable Ising couplings \cite{Islam2013, Monz2016, Debnath2016}. In this system, the long-range spin-spin interaction is mediated by the collective motion of an ion chain and emerges over time scales longer than the propagation time of mechanical waves or phonons through the crystal \cite{Zhu2006a, Zhu2006}. On the other hand, at shorter timescales, such a chain represents a bosonic system of phonon modes that describe the local motion of individual ions. Here each local mode is defined by the harmonic confinement of a particular ion with all other ions pinned. In this picture, phonons hop between the local modes due to the long-range Coulomb interaction between ions \cite{Porras2004, Ivanov2009, Mering2009}. This intrinsic hopping in trapped ion crystals makes it a viable candidate for simulating many-body systems of bosons \cite{Porras2004, Ivanov2009}, boson interference \cite{Toyoda2015} and applications such as boson sampling \cite{Shen2014}.

\FigureOne

Such a system of local oscillators can be approximated to the lowest order of the transverse ion displacement by the phonon Hamiltonian ($\hbar=1$),
\begin{equation}
H_p = \sum_{j}(\omega_{x}+\omega_j) a_j^{\dagger}a_j +\sum_{j<k}\kappa_{jk} (a_j^{\dagger}a_k + a_j a_k^{\dagger}).
\label{Eq1}
\end{equation} 
Here the local mode frequency of each ion is expressed as a sum of the common mode transverse trap frequency $\omega_{x}$ and a position-dependent frequency shift $\omega_j$ experienced by the $j-$th ion \cite{Porras2004,Ivanov2009}. The local mode bosonic creation and annihilation operators are $a_j^{\dagger}$ and $a_j$, respectively. The long-range hopping term $\kappa_{jk} = e^2/(2M\omega_xd_{jk}^3)$ is determined by the distance $d_{jk}$ between ions $j$ and $k$, where $e$ and $M$ are the charge and mass of a single ion. 

By applying external controls to the system, on-site interactions between phonons lead to the simulation of Hubbard models of bosons. For instance, applied position-dependent Stark shifts can result in effective phonon-phonon interactions \cite{Porras2004}. Combined with phonon hopping between sites, such a system follows the Bose-Hubbard model. In the approach considered here, the internal spin is coupled to the external phonon mode by driving the spin resonance on a motion-induced sideband transition \cite{Leibfried2003}. This gives rise to nonlinear on-site interactions between spin-phonon excitations (polaritons). Such a system simulates the Jaynes-Cummings-Hubbard model, which describes an array of coupled cavities \cite{Ivanov2009, Mering2009, Greentree2006, Hartmann2006, Toyoda2013}.

In order to study the dynamics of such bosonic systems, the local phonon modes must be manipulated and detected faster than the hopping rate. Addressing these modes requires fields that target individual ions in space and each local mode in frequency. Previous experiments have observed hopping only between two sites by either using the collective motion of ions in separate but nearby trapping zones \cite{Brown2011, Harlander2011} or by varying the spacing between ions in the same trap to spatially resolve each site \cite{Haze2012}. In contrast, the direct addressing of each local mode in a single ion crystal circumvents such overheads and provides a complete toolbox for implementing larger bosonic systems by simply increasing the number of ions.

In this Letter, we report the observation of free phonon hopping in an ion chain, and study its suppression by applying targeted phonon blockades on individual sites. We access all local motional modes along the transverse direction of a static linear chain of \Yb ions. Phonons are prepared and measured by driving sideband transitions on each mode faster than the rate of hopping in the chain, with an overall fidelity of $89(2)\:\%$. This is achieved by setting the transverse confinement to be much larger than the axial confinement of the chain, which results in suitably low hopping rates due to both a relatively large inter-ion distance ($\sim 10\:\mu$m) and high mass of the \Yb ion.

A phonon blockade is implemented by resonantly driving red-sideband transitions on each site, which couple the internal spin of individually addressed ions to their local phonon mode via a Jaynes-Cummings interaction. In the rotating frame of the free spin and the transverse motional common-mode Hamiltonian ($H_0=\omega_{HF}\sum_j\ket{e}_j\bra{e}_j+\omega_x\sum_ja^{\dagger}_ja_j$), this interaction is represented by the blockade Hamiltonian as
\begin{equation}
H_b=\sum_{j}\Delta_j\ket{e}_j\bra{e}_j+\sum_{j}\frac{\Omega_j^r}{2}(\sigma^+_j a_j  +\sigma^-_j a^{\dagger}_j).
\label{Eq2}
\end{equation}
Here, the spin-$1/2$ `ground' and `excited' states of the $j-$th ion are represented by $\ket{g}_j$ and $\ket{e}_j$, respectively, with energy splitting $\omega_{HF}$, and spin raising and lowering operators $\sigma^+_j$ and $\sigma^-_j$. A local motional red sideband is driven at a Rabi frequency $\Omega_j^r$ and detuned from resonance by $\Delta_j$.

Phonon blockades are applied on individual sites that have ions prepared in the ground state of spin and motion $\ket{g,0}$, where the second index denotes the local mode phonon number. Upon applying the Jaynes-Cummings interaction at resonance ($\Delta_j=0$), a maximal energy splitting of $|\omega_j\pm\Omega^r_j/2|$ occurs between $\ket{g,0}$ and the next excited polaritonic states $\ket{\pm, 1}$. This energy cost suppresses phonons from entering the targeted sites and thereby creates a blockade (see Fig. \ref{Fig1}a). This scheme is analogous to implementing photon blockades using single-atom cavity QED systems \cite{Hamsen2017}.

The experiment consists of a linear chain of three \Yb ions, each with an internal spin defined by a pair of hyperfine `clock' states as $\ket{g}=\ket{F=0,m_F=0}$ and $\ket{e}=\ket{F=1,m_F=0}$ of the $2S_{1/2}$ electronic ground level with a hyperfine energy splitting of $\omega_{HF}=2\pi\times12.642812\:$GHz \cite{Olmschenk2007}. Here, $F$ and $m_F$ denote the quantum numbers associated with the total atomic angular momentum and its projection along the quantization axis defined by an applied magnetic field of $5.2\:$G. The external motion of the trapped ions is defined by a linear rf-Paul trap with transverse (X,Y) and axial (Z) harmonic confinement at frequencies $\{ \omega_x,\omega_y,\omega_z \}=2\pi\times\{3.10,2.85,0.15\}\:$MHz such that the ion chain is aligned along Z with a distance of $d_{j,j+1}=10.1(2)\:\mu$m between adjacent ions. During an experiment, we excite local phonons in the transverse modes along X, which can then hop between the ion sites. The inherent hopping rates are approximately $\kappa_{j,j+1}\approx2\pi\times3\:$kHz and $\kappa_{j,j+2}\approx\kappa_{j,j+1}/8$, respectively. The combined effect of the transverse (X) harmonic confinement and repulsion between ions (determined by $d_{jk}$) define the position-dependent local mode frequency shifts $\{\omega_j\}$. Fig. \ref{Fig1}a represents the local modes with frequencies $\{\omega_j\}$ in a frame rotating at the common mode frequency $\omega_x$. 

\FigureTwo

Coherent control of the spin and motion of each ion is implemented with stimulated Raman transitions using a $355\:$nm mode-locked laser \cite{Hayes2010}, where pairs of Raman beams couple the spin of an ion to its transverse motion \cite{Debnath2016}. A global beam illuminates the entire chain, and a counterpropagating array of individual addressing beams is focused to a waist of $\approx 1\:\mu$m at each ion. The beat note between the Raman beams can then be tuned to $\omega_{HF}$ to implement a ``carrier" transition for coherent spin flips, or tuned to $\omega_{HF}\pm(\omega_x+\omega_j)$ to drive a blue- or red-sideband transition involving local phonon modes. The individual addressing beams are modulated independently using a multi-channel acousto-optic modulator \footnote{Model H-601 Series 32-Channel UV Acousto-Optic Modulator, PN: 66948-226460-G01, Harris Corporation}, each channel of which is driven by a separate arbitrary waveform generator \footnote{Model WX1284C-1 1.25 GS/s Four Channel Arbitrary Waveform Generator, PN: 126182, Tabor Electronics Ltd.}. The wave vector difference $\Delta\bar{k}$ between Raman beams has a projection along both the X and Y directions of motion. Each transverse mode can then be addressed by tuning near their sideband transitions. In order to spectrally resolve each local mode, we choose sideband Rabi frequencies $\Omega^r_j,\Omega^b_j<|\omega_x-\omega_y|$, while also satisfying $|\omega_j|\ll |\omega_x-\omega_y|$ to prevent crosstalk between the modes.

A typical experimental sequence, as shown in Fig. \ref{Fig1}b, starts with the preparation of each ion in state $\ket{g,0}$ by Doppler cooling and subsequent Raman sideband cooling of each of the transverse modes. A single phonon excitation is introduced at a single site by resonantly driving a blue-sideband and carrier $\pi-$pulse to prepare the state $\ket{g,1}$. In order to minimize the effect of hopping during this process, the sideband and carrier $\pi-$pulses are kept short ($\approx10\:\mu$s and $\approx 1\:\mu$s, respectively). Phonon blockades are applied to particular ions, initially prepared in the $\ket{g,0}$ state, by resonantly driving the red-sidebands of their respective local modes. Finally, the single phonon occupancy denoted by states $\ket{g,0}$ and $\ket{g,1}$ is measured at each site using a red-sideband $\pi-$pulse on each ion, which coherently projects it to spin states $\ket{g}$ and $\ket{e}$, respectively. The spin-dependent fluorescence can then be detected using a multi-channel photomultiplier tube, thereby measuring a binary phonon occupancy of 0 or 1 for each site \cite{Olmschenk2007, Debnath2016}.

\TableOne
Figure \ref{Fig2} shows the hopping dynamics. During free hopping, a single excitation is observed to hop predominantly to the neighboring site. The extent of hopping is indicated by the amplitude of the oscillations in phonon occupancy. This is determined by the strength of hopping $\kappa_{jk}$ relative to the energy splitting between local modes $\omega_{jk}=\omega_j-\omega_k$. We observe different hopping rates between ions 1 and 2 compared to that between 2 and 3, which indicates an asymmetry in the local mode energy differences, $|\omega_{12}|\ne|\omega_{23}|$. This is likely due to a stable non-linearity in the transverse confinement of the ion trap. We also note that the sign of the local mode energy difference is critical in governing next-nearest neighbor hopping in systems with three or more modes. This is due to a Raman-type hopping process where appropriate energy splittings between the local modes can facilitate hopping between ion 1 and 3 via ion 2 (see supplementary material).

Phonon hopping is also observed in the presence of a blockade applied on neighboring sites (Fig.\ref{Fig2}d-g). Here, we resonantly drive on the red sideband, creating a ladder of Jaynes-Cummings eigenstates $\{\ket{g,0},\ket{\pm,1},\ket{\pm,2}, ...\}$, where $\ket{\pm,n}$ is a spin-phonon dressed state with polaritonic excitation number $n$ (Fig.\ref{Fig1}a). Since the blockade ion is initially in eigenstate $\ket{g,0}$, hopping into this site is suppressed when the energy splitting of the first excited states $\ket{\pm, 1}$ is much larger than the hopping rate. This implies that when $|\kappa_{jk}|\ll |\Omega_k^r/2\pm\omega_{jk}|$, hopping is suppressed from the $j-$th to $k-$th site, where the tunable blockade strength is set by the red-sideband Rabi frequency $\Omega_k^r$ at the blockaded site $k$. For ions $2$ and $3$, a higher suppression is observed compared to ions $1$ and $2$, where the large phonon mode splitting $\omega_{12}$ results in some residual hopping despite the applied blockade (see supplementary material). 

In Fig.\ref{Fig2}, we further observe nonzero phonon occupancies at sites prepared in the motional ground state owing to imperfect initial sideband cooling. Based on the non-oscillatory near-zero phonon occupancies of the hopping data (Fig.\ref{Fig2} a, c, and g for ions $3$, $1$, and $2$, respectively), we estimate average local mode phonon numbers of $\{\bar{n}_1,\bar{n}_2,\bar{n}_3\}=\{0.09,0.08,0.04\}$. This also leads to an imperfect preparation and measurement of state $\ket{g,1}$, which additionally suffers from residual phonon hopping over the finite duration of sideband $\pi$-pulses.
\FigureThree

We fit the hopping data to a theoretical model using the Hamiltonian presented in Eq.\ref{Eq1} and Eq.\ref{Eq2}, with $\omega_{jk}, \kappa_{jk} $, and $\Omega_{j}^{r}$ as free parameters. The steady-state phonon occupancy, in the absence of hopping (see Fig.\ref{Fig3}), is used to characterize the systematic errors in the preparation and measurement of the states $\ket{g,0}$ and $\ket{g,1}$.  This is subsequently incorporated into all theoretical curves in Fig.\ref{Fig2} \cite{Shen2012} (see supplementary material). The effect of imperfect sideband cooling is directly included by starting from a thermal phonon distribution with mean phonon numbers $\{\bar{n}_j\}$. Both the spin and the motional degrees of freedom are considered in the theoretical simulation of the full hopping dynamics. A single set of parameters (see Table \ref{Tab1}) is used to fit all data in Fig.\ref{Fig2}, which shows a very good agreement between experimental data and the theoretical description of the system. The hopping and blockade strengths given by the parameter values of $\kappa_{jk}$ and $\Omega_j^r$ are consistent with those measured directly from the inter-ion distance and red-sideband spectroscopy, respectively (see supplementary Fig.\ref{FigS1}b).

A straightforward way to improve fidelities of local phonon operations is setting the sideband Rabi frequency much larger than the hopping strength. In this experiment, since the Raman transition drives both transverse modes, the maximum Rabi frequency is limited by the spacing between the modes $(\omega_x-\omega_y = 2\pi\times 250)\:$kHz. In future experiments, this can be resolved by rotating the two principle axes of transverse motion (X and Y) with respect to the Raman beams such that only one mode is excited. For multi-phonon hopping experiments, it is also necessary to implement fast measurement of phonon distributions (instead of single phonon occupancy) using cascaded sideband pulses, as has been implemented on single ions \cite{An2014, Um2016}. 

Tunable local phonon blockades can be a useful tool in studying energy transport in ion chains \cite{Ramm2014}. It can also, in principle, be used in mechanically isolating pairs of ions in a crystal for the implementation of fast entangling gates mediated by their local phonon modes \cite{Zhu2006, Wong-Campos2017}.

This work was supported by the ARO with funds from the IARPA LogiQ program, the AFOSR MURI program on Quantum Measurement and Verification, the ARO MURI program on Modular Quantum Circuits, and the NSF Physics Frontier Center at JQI.

\bibliographystyle{ieeetr}
\bibliography{PhononBlockadeRefs}

\setcounter{figure}{0}
\renewcommand{\thefigure}{S\arabic{figure}}
\setcounter{equation}{0}
\renewcommand{\theequation}{S\arabic{equation}}

\section{Supplementary Materials}

\subsection{1. Spectroscopy of Local Phonon Modes.}

Figure \ref{FigS1} shows the spectroscopy of the motional sidebands of local modes of each ion in a chain of three. The blue sideband spectrum (Fig.\ref{FigS1}a) shows the two transverse motional modes of each ion separated by $\omega_x-\omega_y=2\pi\times250\:$kHz. We choose to excite the $X-$mode, which has the higher mode frequency. The red sideband spectrum is obtained by preparing each ion in the $\ket{g,1}$ state with a single phonon excited in the $X-$mode followed by a red sideband $\pi-$pulse and fluorescence detection (Fig.\ref{FigS1}b). The red sideband Rabi frequencies $\Omega_j^r$ of the local modes are extracted from the spectroscopy and are shown in Table \ref{Tab1}.

The transverse common mode trap frequencies ($\omega_x$, $\omega_y$) are set by the amplitude of the ion trap rf signal. This can drift over time causing the local mode frequencies to vary. Therefore, in order to mitigate errors in the spectral addressing of each local mode, the transverse confinement $(\omega_x)$ is actively stabilized \cite{Johnson2016}. 

\subsection{2. State Preparation and Measurement (SPAM) Error Correction. }
In the experiment, the state preparation and measurement are not perfect, due to single ion fluorescence detection error, imperfect cooling ($\bar{n} > 0$), and mapping of phonon occupancy. For detection, the phonon number states $\ket{0}$ and $\ket{1}$ are mapped onto the spin states $\ket{g}$ and $\ket{e}$ respectively. 

We consider two separate single-ion SPAM errors, $\epsilon_{g}$ for $\ket{g}$ and $\epsilon_{e}$ for $\ket{e}$. In order to determine these, we prepare the ion in state $\ket{g}$ and $\ket{e}$ and measure the fraction of bright events. Our readout errors are $\epsilon_{g} = 0.26\%$ and $\epsilon_{e} = 0.91\%$. The experimental readout $P_{g} = 1-P_{e}$ and $P_{e}$ are then related to theoretical predictions by \cite{Shen2012}
\begin{equation}
\left(\begin{array}{c}P_{g} \\P_{e}\end{array}\right) = 
\left(\begin{array}{cc}1-\epsilon_{g} & \epsilon_{e} \\ \epsilon_{g} & 1-\epsilon_{e}\end{array}\right)
\left(\begin{array}{c} P_{g}^{\text{th}}  \\ P_{e}^{\text{th}}  \end{array}\right). 
\end{equation}

We now include the effect of imperfect cooling. After sideband cooling, the phonon distribution is assumed to be thermal with an average phonon number of $\bar{n}$. Then, we can consider the initial state to be an incoherent mixture of different phonon numbers, with probabilities $P_{n} = \bar{n}^{n}/(\bar{n}+1)^{n+1}$. Hence, the effect of imperfect cooling can be directly incorporated in the theoretical model.

Finally, we extend the error model to include another layer of SPAM error in the detection of phonon occupancy due to imperfect mapping to spin states with errors $\epsilon_{g}'$ and $\epsilon_{e}'$. Assuming the ground state preparation is limited only by fluorescence detection error and the effect of the thermal phonon distribution ($\bar{n}>0$), i.e., $\epsilon_{g}' = 0$, we have
\begin{equation}
\left(\begin{array}{c}P_{g} \\P_{e}\end{array}\right) = 
\left(\begin{array}{cc}1-\epsilon_{g} & \epsilon_{e} \\ \epsilon_{g} & 1-\epsilon_{e}\end{array}\right)
\left(\begin{array}{cc}1 & \epsilon_{e}' \\0 & 1-\epsilon_{e}'\end{array}\right) 
\left(\begin{array}{c} P_{g}^{\text{th}}(\bar{n})  \\ P_{e}^{\text{th}}(\bar{n})  \end{array}\right), 
\end{equation}
where the theoretical values $P_{g}^{\text{th}}(\bar{n})$ and $P_{e}^{\text{th}}(\bar{n})$ depend on the mean phonon number $\bar{n}$. From Fig.\ 3 of the main text, the experimentally-measured probabilities are $P_{e} \approx 0.836$ and $P_{e} \approx 0.048$ for the state $\ket{g,1}$ and $\ket{g,0}$, respectively (averaged over time and ions). Fitting the theoretical values to those, we have two variables, $\bar{n}$ and  $\epsilon_{e}' $, and two equations, from which we can extract the values as $\bar{n} \approx 0.055$ and $\epsilon_{e}' \approx 0.114$. The SPAM error matrix is
\begin{equation}
\begin{split}
M_{\text{SPAM}} &= 
\left(\begin{array}{cc}1-\epsilon_{g} & \epsilon_{e} \\ \epsilon_{g} & 1-\epsilon_{e}\end{array}\right)
\left(\begin{array}{cc}1 & \epsilon_{e}' \\0 & 1-\epsilon_{e}'\end{array}\right) \\
&\approx \left(\begin{array}{cc}0.997 & 0.122 \\0.003 & 0.878 \end{array}\right),
\end{split}
\end{equation}
which is then used to scale all theoretical curves.
\\
\subsection{3. Long-range Hopping and Raman-like Hopping. }
The phonon hopping amplitude $\kappa_{ij}$ decays as $1/d_{ij}^{3}$, where $d_{ij}$ is the distance between ions $i$ and $j$, so direct long-range hopping $\kappa_{13} \approx \kappa_{12}/8$. Hence, we expect the effect of long-range hopping in the three ion chain to be unimportant. A natural approach is to use the Rabi flopping of the free hopping data in Fig.\ref{Fig2}a of the main text to extract the local mode energy difference $\omega_{12}$ and $\kappa_{12}$. Similarly, $\omega_{23}$ and $\kappa_{23}$ are expected to be extracted in the same way from Fig.\ref{Fig2}c. However, this picture of two-site (two-level) approximation is incomplete, primarily due to a Raman-like process of hopping, i.e., phonons hopping from ion 1 to ion 2 and then from ion 2 to ion 3. We have confirmed our observation by using the two-ion approximation to extract out the experimental parameters; the fitted parameters deviate substantially from those in Table I of the main text and the former produce poor fits to other panels of data.

In addition, the relative sign between $\omega_{12}$ and $\omega_{23}$ does not matter in the two-ion picture. However, considering the dynamics of the full system of three modes, the relative sign is important. One can again understand why this is the case via an analogy to a Raman process. Phonons can hop from ion 1 to ion 3 via ion 2. If $\omega_{12}$ and $\omega_{23}$ had opposite signs, the magnitude of $\omega_{13}$ would be much smaller (analogous to the two-photon detuning in Raman transitions). This would lead to a Raman-like hopping from ion 1 to 3, which would induce much stronger oscillations than are observed in the experiment.

Therefore, we need to consider phonon hopping among three ions to correctly model their full dynamics. In the theoretical fitting, we use seven free parameters $\omega_{12}, \omega_{23}, \kappa_{12}, \kappa_{23}, \Omega^{r}_{1}, \Omega^{r}_{2}, \Omega^{r}_{3}$ to fit all experimental data in Fig.\ref{Fig2} of the main text, taking into account the SPAM error and the thermal phonon distribution after sideband cooling.  
\\
\subsection{4. Additional Explanation on the Phonon Blockade Data.}
This section provides some additional explanation on a few non-intuitive behaviors of the blockade data. To understand it better, we include a simplified three-level picture. Consider Fig.\ref{Fig2}d of the main text. Three collective states are predominantly involved: $\ket{g,1}_{1}\ket{g,0}_{2}\ket{g,0}_{3}$, $\ket{g,0}_{1}\ket{g,1}_{2}\ket{g,0}_{3}$, and $\ket{g,0}_{1}\ket{e,0}_{2}\ket{g,0}_{3}$, where the subscript labels the ion number. In this subspace, the Hamiltonian can be approximated as 
\begin{align}
H &= \left(\begin{array}{ccc}
0 & \kappa_{12} & 0 \\
\kappa_{12} & -\omega_{12} & \Omega_{2}^{r}/2 \\
0 & \Omega_{2}^{r}/2  & -\omega_{12}
\end{array}\right) ,
\label{Eq:3level}
\end{align}
where $\kappa_{12}/2\pi \approx 2.9\,$kHz, $\omega_{12}/2\pi \approx 11.58\,$kHz, and $\Omega_{2}^{r} \approx 45.9\,$kHz (the common energy $\omega_{1}$ is subtracted out). Instead of suppressing hopping, the local mode frequency difference $\omega_{12}$ facilitates hopping due to the fact that it partially offsets the blockade strength $\Omega_{2}^{r}/2$. Figures \ref{FigS2}a and \ref{FigS2}b compare the scenarios in which $\omega_{12}/2\pi = 11.58\,$kHz and $\omega_{12}/2\pi = 0\,$kHz, respectively. It is evident that hopping is further suppressed for the case $\omega_{12}/2\pi = 0\,$kHz. Since $|\omega_{12}| > |\omega_{23}| $, this explains why the blockade data in Fig.\ref{Fig2}d seems to have stronger oscillation than that in Fig.\ref{Fig2}g. In addition, from Fig. \ref{FigS2}, we also see that much of the state population is stored in $\ket{g,0}_{1}\ket{e,0}_{2}\ket{g,0}_{3}$, which explains the observation that, for Fig. \ref{Fig2}d, ion 1 seems to have a stronger oscillation than ion 2; the apparent non-conservation of the total phonon numbers is due to our detection method of mapping phonons to spins via a single red sideband pulse. 

In addition, we can understand how the red sideband Rabi frequencies come into play for the blockade data. For the free hopping data, red sideband Rabi frequency does not play a role other than some small hopping during the short pulse period. So the free hopping data (Fig.\ref{Fig2}a-c) is not sensitive to $\Omega_{1}^{r}, \Omega_{2}^{r}, \Omega_{3}^{r}$. However, for the blockade data (Fig.\ref{Fig2}d-g), due to a much longer hopping period and the offset between local mode frequency difference and the blockade strength, the observations depend sensitively on  $\Omega_{1}^{r}, \Omega_{2}^{r}, \Omega_{3}^{r}$. Therefore, we also fit the red sideband blockade strengths as free parameters in the theoretical model.

\FigureFour
\FigureFive

\end{document}